\documentclass[preprint2,times,tighten]{aastex6}

\usepackage{graphicx}
\usepackage{dcolumn}
\usepackage{bm}
\usepackage{hyperref}
\usepackage{url}
\usepackage{natbib}

\usepackage{xcolor}

\usepackage{soul}

\usepackage{multirow}

\newcommand\lb{\langle}              
\newcommand\rb{\rangle}               

\begin{document}
	

\title{Sustained Turbulence in Differentially Rotating Magnetized Fluids at Low Magnetic Prandtl Number}

\author{Farrukh Nauman and Martin E. Pessah}
\email{nauman@nbi.ku.dk}
\affil{Niels Bohr International Academy, The Niels Bohr Institute, Blegdamsvej 17, DK-2100, Copenhagen \O, Denmark.}
	

\date{\today}




\begin{abstract}
We show for the first time that sustained turbulence is possible at low magnetic Prandtl number for Keplerian flows with no mean magnetic flux. Our results indicate that increasing the vertical domain size is equivalent to increasing the dynamical range between the energy injection scale and the dissipative scale. This has important implications for a large variety of differentially rotating systems with low magnetic Prandtl number such as protostellar disks and laboratory experiments.
\end{abstract}

	
\maketitle


\section{Introduction}  
Differentially rotating flows are ubiquitous in nature, from accretion flows around young stars and compact objects to convection zones inside stars. Magnetic fields are important in many of these flows. Understanding the stability properties of these flows can shed light into the processes driving the transport of angular momentum and energy in these environments. The Magnetorotational Instability (MRI) is a linear instability that has emerged as a potential explanation for the origin of turbulence in weakly magnetized differentially rotating flows (e.g. magnetized Taylor-Couette flows, accretion disks \cite{1959velikhov, 1960PNAS...46..253C, 1991ApJ...376..214B}). MRI requires a weak magnetic flux to work but given the paucity of direct observations of magnetic field amplitudes and geometries, it is of interest to explore a range of cases for the magnetic flux threading an accretion flow, including the case of zero mean magnetic flux. 


The earliest 3D simulation of ideal magnetohydrodynamic (MHD) Keplerian shear flow without a mean magnetic flux by  \cite{hawley1996} showed that the saturated level of turbulent stresses decreases with resolution (see also \cite{pessah2007}). Later work by  \cite{fromang2007} used physical diffusion coefficients and showed that turbulence does not persist for magnetic Prandtl number, $Pm = Rm/Re \lesssim 2$. Recent work on linearly stable hydrodynamic shear flows (e.g., \cite{hof2006}) suggests that turbulence lifetime is finite and increases exponentially with the Reynolds number, $Re$. \cite{rempel} explored the question of turbulence lifetime in Keplerian shear flows with zero net magnetic flux with a small vertical domain size ($L_z=1$) and found that the turbulence lifetime increases exponentially with $Rm$ between $9,000$ and $11,000$ ($Re=3,125$). 

Examples of $Pm \ll 1$ systems include astrophysical systems such as protostellar disks (for more astrophysical examples, see table 1 of  \cite{brandenburg2005}) and laboratory plasmas (e.g., \cite{sisan2004}, \cite{exp2014}). The $Pm \ll 1$ limit has been studied extensively in non-helically forced isotropic MHD turbulence simulations and theory, with the consensus that a larger $Rm_{\text{crit}}$ is required in this limit compared to the $Pm \gg 1$ \cite{sch2007}. Earlier work by  \cite{sch2004} had claimed that the small scale dynamo (growth of the magnetic field on dissipation scales) does not exist in the low $Pm$ limit but later work \citep{iskakov2007} demonstrated growth and sustenance of magnetic fields (at a higher $Rm_{\text{crit}}$). In their study,  \cite{sch2004} conjectured that a low $Pm$ dynamo might only be possible in the presence of a mean field. Indeed for MRI simulations with net magnetic flux, turbulence can be sustained in the low $Pm$ regime (\cite{lesur2007}, \cite{meheut2015}). 

The question of whether a large domain size can have a significant effect on flow stability has been long under discussion (e.g., \cite{pomeau1986}, \cite{philip2011}). The shearing box is a local approximation \cite{1965MNRAS.130...97G} for differentially rotating flows such as accretion disks (or Taylor-Couette flows) and it is unclear whether it can exhibit similar behavior as the spatiotemporal chaotic patterns that might emerge in a realistic accretion flow with its very large spatial extent and very large $Re$ (\cite{cross1993}). Nevertheless, within the shearing box framework, it is of interest to explore how a small domain size (`minimal flow unit' \cite{jimenez1991}, \cite{rincon2007}) is different from a system with larger degrees of freedom as a result of larger domain size, resolution, $Re$ and $Rm$. Using an asymptotic analysis with net toroidal and vertical flux, \cite{julien2007} showed that scale separation between the most unstable mode and the vertical domain size leads to a saturated state of MRI where the energy is dominated by the box scale in the vertical direction.

In this Letter, we explore the question of whether sustained turbulence can be found in unstratified\footnote{ \cite{davis2010} suggest that density stratification due to gravity lowers the critical $Pm$ slightly but they did not explore the $Pm<1$ regime.} incompressible MHD Keplerian shear flows with zero mean magnetic flux with a variety of box sizes and dissipation coefficients. We find that the turbulence lifetime is very sensitive to the vertical box size \footnote{We do not explore elongated azimuthal domains as \cite{riols2015} already explored $L_y=20$ and did not find evidence for sustained turbulence at $Pm < 1$.} and that large scale structures develop both in the velocity and the magnetic fields. 

\section{Numerical Simulations and Results} 
We use the publicly available pseudospectral code \textsc{snoopy} \footnote{\url{http://ipag.osug.fr/~lesurg/snoopy.html}} (\cite{lesur2007}). All of our simulations employ a zero net flux initial field ${\bm B}_{\text{ini}} = B_0 \sin(k_x x)\, {\bm e}_z$. The shear profile is ${\bm V}_{\text{sh}} = - Sx\, {\bm e}_y$, where $S=q\Omega=1$ ($q=1.5$ for Keplerian shear) is the shear parameter and $\Omega$ is the angular frequency. We apply perturbations of order $L S$ to the first few velocity modes. All of our results are reported in units of shear times $1/S$ (for comparison, $10,000$ shear times are equal to about $1,061$ orbits ($2\pi/\Omega$) in our units). The dissipation coefficients are characterized by the Reynolds number, $Re = SL^2/\nu$ and the magnetic Reynolds number, $Rm = SL^2/\eta$;  where $L=1$. For most of our runs, $L_x = L$. The magnetic Prandtl number $Pm=Rm/Re=\nu/\eta$, is of course independent of the choice of characteristic scale. We report our domain sizes as $L_x\times L_y\times L_z$.

\begin{figure}
	\centering
	\includegraphics[width=0.45\textwidth]{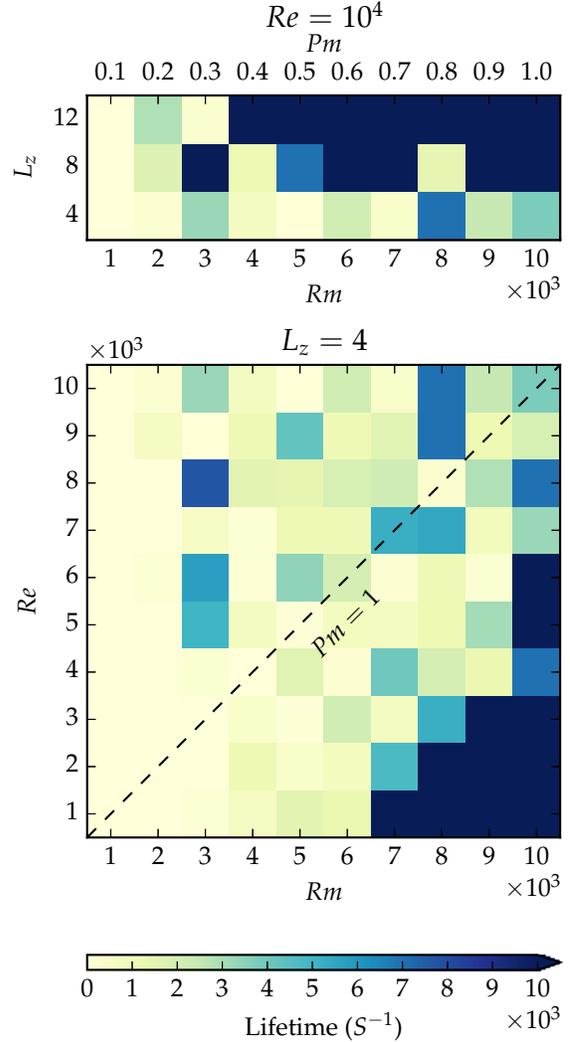}
	
	\caption{Turbulence lifetime in units of $1,000 S^{-1}$ for runs with: (i) fixed $Re=10,000$ but variable vertical box size $L_z$ and $Pm=Rm/Re$  (top panel); (ii)  fixed vertical box size $L_z=4$ ($1\times 2\times 4$) but variable $Re$ and $Rm$ (bottom panel). The lifetime of turbulence increases with larger $Re$ and $Rm$, but it is most sensitive to $L_z$. Note that we stopped all of our simulations at $10,000$ shear times so the runs that show $10,000$ shear times might have a much larger lifetime. The number of zones used for these simulations are $64\times 64\times (64*L_z)$.}
	\label{fig:LzvsPm}
\end{figure}


\subsection{Fixed $Re$, Variable $L_z$ and $Pm$}
In the top panel of fig. \ref{fig:LzvsPm}, we find that as we increase $L_z$ while keeping $L_x=1$ and $L_y=2$, even $Pm<1$ manages to sustain turbulence for several thousand shear times. This is to be contrasted with earlier work of  \cite{fromang2007} ($L_z=1, Re=3,125$) and  \cite{shi2016} ($L_z=4, Re=3,125$), which showed that $Pm \geq 2$ is required for sustained turbulence.  \cite{shi2016} reports that $Pm=2$ simulation sustained turbulence for thousands of shear times while their $Pm=1$ run decayed after about $2,000$ shear times. In an extensive numerical study,  \cite{rempel} (using \textsc{snoopy}) showed that turbulence is transient and the lifetime of turbulence increases exponentially with $Rm$ (for $2.88 \leq Pm \leq 3.52$) with $L_z=1,Re=3,125$. We do not conduct such an exhaustive study here to determine the functional dependence of lifetime on $L_z$ and $Pm$, since a way of conducting such a study is to evolve a flow to a fully turbulent state and then take that state as initial condition for runs with different dissipation coefficients \citep{rempel}. Due to the spatiotemporal chaotic nature of turbulence, it can take hundreds of runs for each $L_z$ to get good statistics.

\subsection{Fixed $L_z$, Variable $Re$ and $Pm$}
We show lifetime of the turbulent flow as a function of $Re$ and $Rm$ with fixed $L_z=4$ in the bottom panel of fig. \ref{fig:LzvsPm}. The lifetime increases with $Pm$ (and $Rm$, the bottom right of the panel). More importantly, we find that at high enough $Re$, turbulence can last several thousand shear times at $Pm<1$. Combined with fig. \ref{fig:LzvsPm}, this result hints that as $Re \rightarrow \infty$ and $L \gg 1$, $Pm_{\text{crit}}$ reaches a low asymptotic value (\cite{fromang2007}) and our work suggests that $Pm_{\text{crit}} < 1$ \citep{sch2004}.

\begin{figure}
	\centering
	\includegraphics[width=0.45\textwidth]{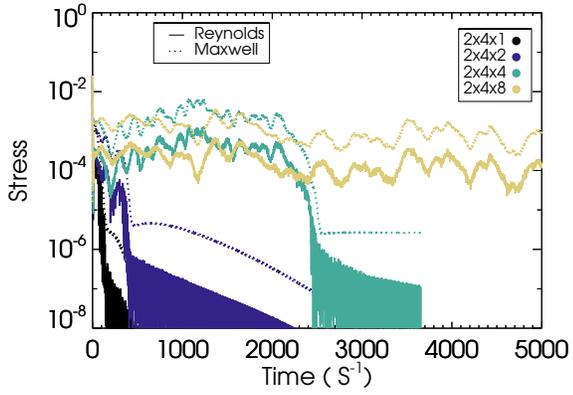}
	
	\caption{Time history for runs with the same horizontal domain size ($L_x=2$, $L_y=4$) as \cite{walker2016} but larger vertical domains. We see sustained turbulence even though our $Re=Rm=10,000$, much lower than that of \cite{walker2016}.}
	\label{fig:2x4xLz}
\end{figure}

\cite{walker2016} did simulations with a domain size of $2\times 4\times 1$ at an unprecedented high resolution of $1024\times 1024\times 512$ where they fixed $Re=45,000$. They found that zero net flux simulation immediately decays at $Pm=1$. We used a lower resolution but larger vertical domain size $L_z$ ranging from $1, 2, 4, 8$ at fixed $Re=Rm=10,000$. As shown in fig. \ref{fig:2x4xLz}, the $2\times 4\times 4$ run is turbulent for $2,000 S^{-1}$ until it decays while the $2\times 4\times 8$ run remains turbulent for the entire duration of our run (and possibly much longer), which is $10,000 S^{-1}$. This clearly demonstrates that vertical domain size is an important parameter in transition to turbulence studies in the shearing box.

\begin{figure}
	\centering
	\includegraphics[width=0.45\textwidth]{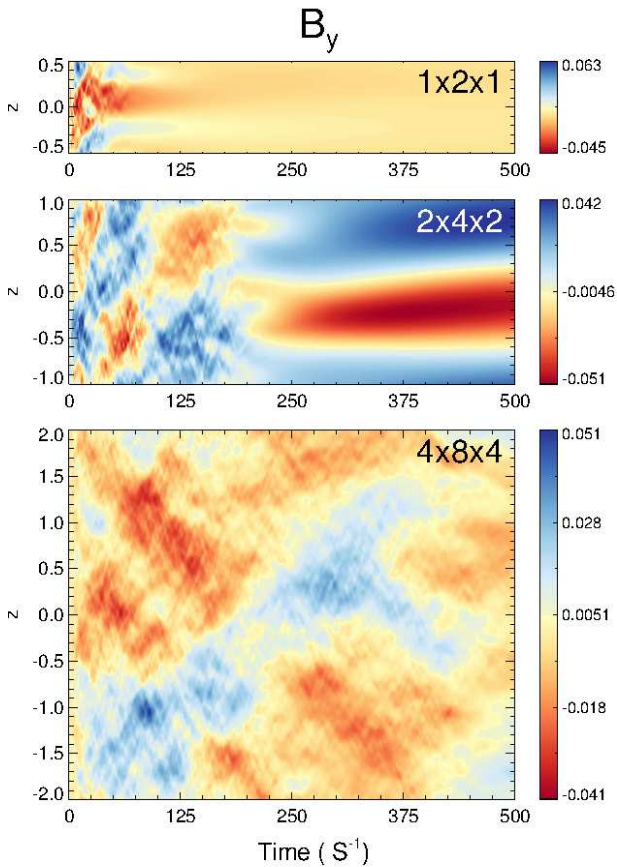}
	
	\caption{Comparison of the $\lb B_y\rb$ (angled brackets represent $xy$ average) profiles for three different box sizes $1\times 2\times 1$, $2\times 4\times 2$, $4\times 8\times 4$ with $Re=Rm=10,000$ with the same aspect ratio. Turbulence is only sustained for significant duration for the largest domain, $4\times 8\times 4$. }
	\label{fig:threeboxes}
\end{figure}

\subsection{Same aspect ratio ($L_z/L_x$=1), larger box size}
We do not find any significant evidence for turbulence in a domain with size $1\times 2\times 1$ (with a resolution of $64^3$) and short lived turbulence ($\sim 175$ shear times) for $2\times 4\times 2$ ($128^3$). The $4\times 8\times 4$ ($256^3$) run is turbulent for nearly $5,000$ thousand shear times. The dissipation coefficients are set by $Re=Rm=10,000$, which is independent of the absolute box size since we scale both coefficients with a fixed $L=1$. We plot the $\lb B_y\rb$ (angled brackets represent $xy$ average) for the three runs $1\times 2\times 1$, $2\times 4\times 2$ and $4\times 8\times 4$ in fig. \ref{fig:threeboxes}. A coherent field develops in all cases but only the biggest box size manages to sustain turbulence for a long time. To compare the effects of increasing the domain size to the effect of increasing the resolution, we ran a simulation at $1\times 2\times 1$ with a resolution of $256^3$ (the same number of zones used in the $4\times 8\times 4$) and found no signs of turbulence with $Re=Rm=10,000$. This comparison suggests that the turbulence lifetime is more sensitive to the box size than the resolution.

\begin{figure}
	\centering
	\includegraphics[scale=0.525]{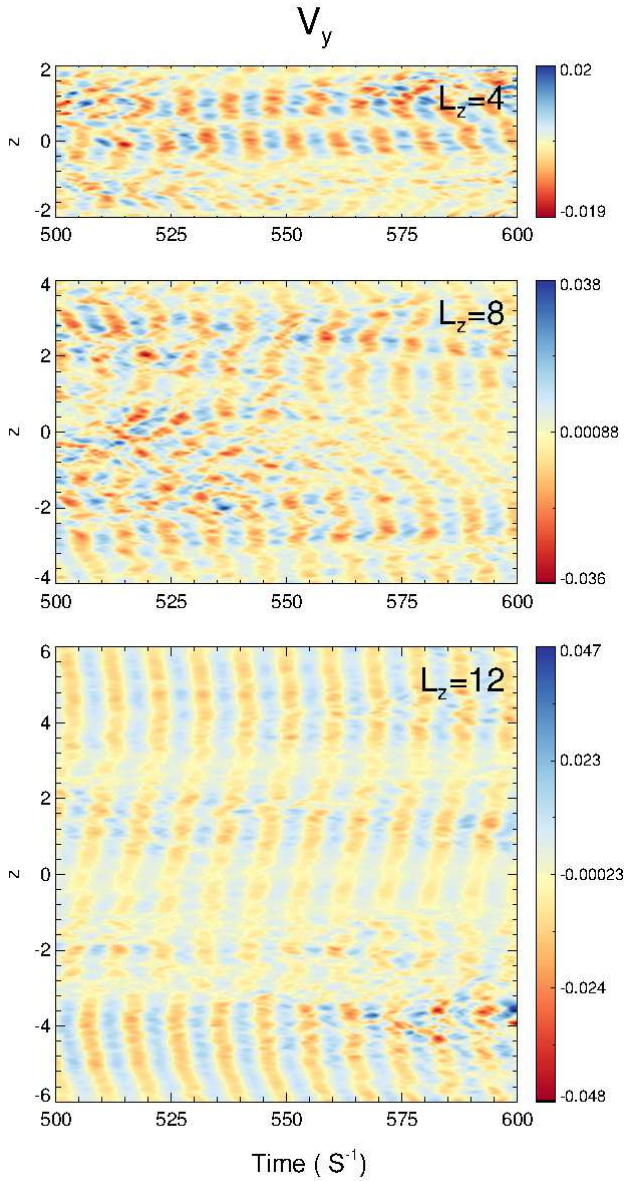}
	
	\caption{Comparison of the $\lb V_y\rb$ profiles for three different vertical box sizes $1\times 2\times L_z$: $L_z=4,8,12$ with $Re=Rm=10,000$. A coherent velocity field ($\sim \sin(\kappa t + \sin k_z z)$), which is dominated by the vertical box scale suggests that the vertical box size has a very significant role in determining the (kinetic) `energy containing' scale of the system.}
	\label{fig:VyLz}
\end{figure}

\subsection{Velocity fields} 
We plot the azimuthal velocity profile $\lb V_y\rb$ in fig. \ref{fig:VyLz} for three different vertical box sizes $L_z=4,8,12$ while keeping $L_x=1, L_y=2, Re=Rm=10,000$. The spatiotemporal dependence of the velocities suggest that velocities behave like: $\sin(\kappa t + \sin k_z z)$, where $k_z = 2\pi n_z/L_z$, and $2\pi/\kappa \sim 9.42 S^{-1}$ is the epicyclic time, with $\kappa^2 =  2\Omega^2(2-q)$. The vertical profile of different velocity components seems to be dominated by $n_z=1$ regardless of box size, which indicates that the gap between the energy containing scale and the dissipation scale increases with the increase in the vertical domain. The corresponding azimuthal magnetic field component profiles for these runs behave much in the same way as in fig. \ref{fig:threeboxes} so they are also dominated by $n_z=1$ irrespective of the domain size. Our results are largely consistent with the recent suggestion by  \cite{jimenez2016}, that shearing box flow has the vertical box size as the outer scale. 

\subsection{Magnetic fields (dynamo)} 
Both  \cite{shi2016} and \cite{walker2016} refer to `dynamo' in their discussion, with the focus on the large scale in the former and small scale in the latter. One might ask which one of these two different mechanisms (or both) is responsible for the growth and sustenance of magnetic fields in a zero net flux MHD Keplerian flows. This distinction is difficult since it is not entirely clear whether there is a separation of \textit{spatial} scales between the box scale and the forcing scale. An argument can be made that since the azimuthal magnetic field varies on several epicyclic times (e.g. $4\times 8\times 4$ run has a cycle period of about $500S^{-1}$, see fig. \ref{fig:threeboxes}) as opposed to one epicyclic time ($\sim 9.42 S^{-1}$) for the velocity field, there is at least a \textit{temporal} scale separation \footnote{We thank Eric Blackman for pointing this out.} and hence a large scale dynamo (\cite{gressel2015}, \cite{bhat2016}). 

The small scale dynamo, on the other hand, refers to the growth of magnetic fields on the viscous scale ($Pm > 1$) or the resistive scale ($Pm < 1$) (\cite{sch2007}) and it might be the mechanism responsible for increasing the lifetime as $Rm$ increases. There is yet another possibility: growth due to a mean velocity flow (e.g., \cite{catt2005}, \cite{mininni2005}). This might be quite significant in the simulations we report here since the velocities seems to settle into large scale periodic structures (fig. \ref{fig:VyLz}).  \cite{sch2007} suggest comparing the growth rate dependence of the magnetic field on $Rm$ to distinguish between the small scale dynamo and the mean velocity flow driven dynamo but since we trigger our simulations with finite amplitude perturbations, it is unclear how to determine the growth rates and hence distinguish between the two.

\subsection{Resolution test} 

\begin{figure}[th]
\centering
\includegraphics[width=0.45\textwidth]{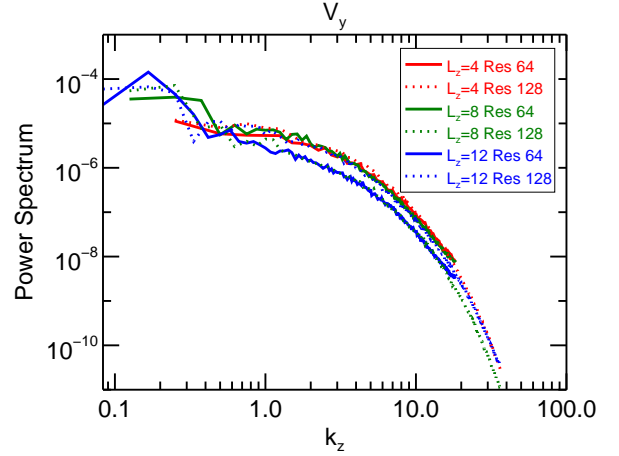}
\caption{Power spectrum of $\langle V_y\rangle (k_z)$ averaged over $251-500 S^{-1}$, where $k_z = 2\pi n_z/L_z$ for $L_z = 4,8,12$ at two different resolutions $64\times 64\times (64*L_z)$ and $128\times 128\times (128*L_z)$. $Re=Rm=10,000$ for all of these runs. This plot suggests that resolution does not significantly alter the azimuthal velocity as a function of resolution.}
\label{fig:spectraVy}
\end{figure}
\begin{figure}[th]
	\centering
	\includegraphics[width=0.45\textwidth]{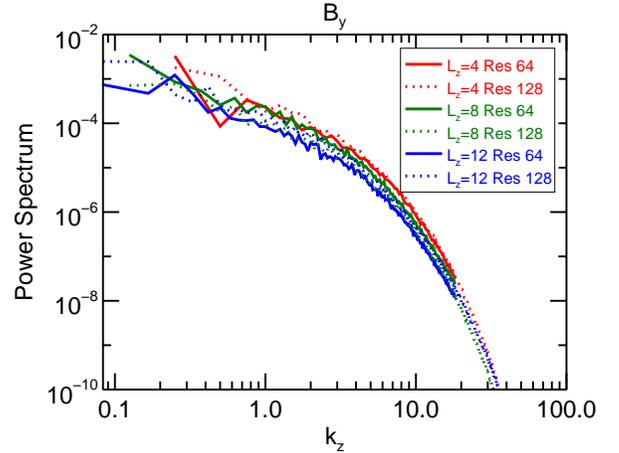}
	\caption{Same as fig. \ref{fig:spectraVy} but power spectrum of $\langle B_y\rangle (k_z)$. }
	\label{fig:spectraBy}
\end{figure}

As a resolution check, we compare the power spectrum in $k_z$ ($= 2\pi n_z/L_z$) of three different runs $1\times 2\times L_z$ at $L_z=4,8,12$ ($Re=Rm=10,000$) at two different resolutions $64\times 64\times (64*L_z)$ and $128\times 128\times (128*L_z)$ in figs. \ref{fig:spectraVy} and \ref{fig:spectraBy}. We find that the increase in resolution does not significantly alter the turbulent power spectrum. There is some discrepancy at low $k_z$ but we point out that since both $V_y$ and $B_y$ show cyclic behavior, the chosen range and initial point for temporal averaging can have a significant effect on the largest wavelengths.

\section{Discussion and Conclusions}
Recent numerical simulations and experiments on hydrodynamic shear flows suggest that transition to turbulence mimics a second order phase transition and can be described by the directed percolation universality class (\cite{shih2016NatPh, lemoult2016NatPh, sano2016NatPh}). The basic idea is that linearly stable flows develop two kinds of domains: laminar and turbulent corresponding to dead and active states. It is through the interaction of these two types of domains that turbulence starts to spread and eventually fills the whole domain as the Reynolds number is increased. This might explain why the box size plays a special role: a larger box size allows for more turbulent and laminar domains to fit in the box and thus allows for complex pattern formation (\cite{cross1993}). A potential analogy for our simulations is the generation of $B_x$ (active state) through stochastic processes on small scales and the subsequent generation of $B_y$ (dead state) through the $S B_x$ term (see also the recent study by \cite{riols2016} who describe the dynamo using the concept of `active' and `slave' perturbations). 

We have demonstrated for the first time that Keplerian shear MHD turbulence (without using any external forcing or net magnetic flux) can sustain for several thousand shear times at $Pm \leq 1$ (for $Re=10,000$), if one uses domain sizes with $L_z \geq 4$. We find that with the $Re$ and $Rm$ up to $10,000$ there is no turbulence at $Pm \leq 1$ when $L_z = 1$, consistent with previous work. It seems that by increasing $L_z$, one increases the separation between the outer scale ($\sim L_z$) and the dissipation scale, which is qualitatively similar to the effect of increasing $Rm$. But the increase in $L_z$ at fixed $Re$ and $Rm$ has a more dramatic effect on turbulence lifetime than increasing $Re$ and $Rm$ at $L_z=1$. Our work clearly illustrates that flow stability studies should not be confined to very small domains and has especially important implications for laboratory plasmas, which can only explore the $Pm\ll 1$ regime {\cite{sisan2004}, \cite{exp2014}}. 

\acknowledgments{ We thank Shantanu Agarwal, Eric Blackman, Oliver Gressel, Tobias Heinemann, Paul Manneville, Jiming Shi, Jim Stone and Francois Rincon for discussions. Colin McNally is thanked for suggestions and help with the figures. The computations were performed on the BlueStreak cluster at the Center for Integrated Research Computing (CIRC) at the University of Rochester. The research leading to these results has received funding from the European Research Council under the European Union’s Seventh Framework Programme (FP/2007-2013) under ERC grant agreement 306614.}

\bibliography{general}
\bibliographystyle{apj}

\end{document}